\documentclass[prd,showpacs]{revtex4}
\begin{document}

\newcommand{\re}{\mathop{\mathrm{Re}}}

\newcommand{\be}{\begin{equation}}
\newcommand{\ee}{\end{equation}}
\newcommand{\bea}{\begin{eqnarray}}
\newcommand{\eea}{\end{eqnarray}}
\addtolength{\topmargin}{25mm}

\title{Canonical superenergy and angular supermomentum complexes in general relativity
and some of their applications}

\author{Janusz Garecki}
\email{garecki@wmf.univ.szczecin.pl}
\affiliation{\it Institute of Mathematics University of Szczecin
and Cosmology Group University of Szczecin,
 Wielkopolska 15, 70-451 Szczecin, Poland}
\date{\today}

\input epsf

\begin{abstract}
Many years ago we have introduced into general relativity, {\bf GR}, the canonical superenergy tensors, $S_i^{~k}$, and the canonical
angular supermomentum tensors, $S^{ikl}=(-)S^{kil}$, matter and
gravitation. We have obtained these tensors  by special averaging of the differences of the
canonical energy-momentum and canonical angular momentum. The
averaging was performed in Riemann normal coordinates, {\bf RNC(P)}; {\bf
P} is beginning of these coordinates.

About four years ago we have observed that these tensors can also
be obtained in other, simpler way, by using the canonical
superenergy and angular super momentum complexes, $_K  S_i^{~k}$,
and, $_K S~^{ikl}=(-)_K S^{kil}$, respectively.
Such complexes can be introduced into {\bf GR} in a natural way
starting from canonical energy-momentum and angular momentum
complexes.

In this paper, at first, we define the canonical superenergy and angular
supermomentum complexes in {\bf GR} and then, we apply them to
analyze of a closed system, {\bf CS}, Trautman's radiative
spacetimes, {\bf TRS}, and Friedman universes, {\bf FU}.

Finally, we compare these complexes and the results obtained with their
help with the canonical superenergy and angular supermomentum
tensors and results obtained with them in past.

In Appendix, for convenience, we summarize our old approach to
canonical superenergy and angular supermomentum tensors.
\end{abstract}


\maketitle
\section{The canonical superenergy and angular supermomentum complexes in {\bf
GR}}
We begin with general remark that in the whole paper we will use
the same signature $(+---$ and notation as used in the last
edition of the famous book \cite {LL}. Latin indices take values $0,1,2,3$
and Greek indices range values $1,2,3$.

In the framework  of general relativity, {\bf GR}, as a consequence of the Einstein Equivalence
Principle, {\bf EEP}, the gravitational field {\it has non-tensorial strengths} $\Gamma^i_{kl}
 = \{^i_{kl}\}$ and {\it admits no energy-momentum tensor}. By using standard procedures of the
classical field theory one can only attribute to this field {\it gravitational
 energy-momentum pseudotensors}. The leading object of such a kind
 is the {\it canonical gravitational energy-momentum pseudotensor}
 ,$_E t_i^{~k}$, proposed already in past by Einstein. This
 pseudotensor is a part of the {\it canonical energy-momentum
 complex} ,$_E K_i^{~k}$, in {\bf GR}.

The canonical complex, $_E K_i^{~k}$, firstly obtained by using standard field-theoretic
procedure to general relativistic Lagrangian (See, e.g., \cite{Tol, Web, LL}), can be most
easily obtained by rewriting Einstein equations to the superpotential form
\begin{equation}
_E K_i^{~k} := \sqrt{\vert g\vert}\bigl( T_i^{~k} + _E
t_i^{~k}\bigr) = _F U_i^{~[kl]}{}_{,l},
\end{equation}
where $T^{ik} = T^{ki}$ is the symmetric energy-momentum tensor for matter, $g = det[g_{ik}]$,
 and
\begin{eqnarray}
_E t_i^{~k}& =& {c^4\over 16\pi G} \bigl\{\delta_i^k
g^{ms}\bigl(\Gamma^l_{mr}\Gamma^r_{sl} -
\Gamma^r_{ms}\Gamma^l_{rl}\bigr)\nonumber\\
&+& g^{ms}_{~~,i}\bigl[\Gamma^k_{ms} - {1\over 2}
\bigl(\Gamma^k_{tp}g^{tp} -
\Gamma^l_{tl}g^{kt}\bigr)g_{ms}\nonumber\\
&-& {1\over 2}\bigl(\delta^k_s \Gamma^l_{ml} +
\delta^k_m \Gamma^l_{sl}\bigr)\bigr]\bigr\};
\end{eqnarray}
\begin{equation}
_F {U_i^{~[kl]}} = {c^4\over 16\pi G}{g_{ia}\over\sqrt{\vert
g\vert}}\bigl[\bigl(-g\bigr)\bigl(g^{ka} g^{lb} - g^{la}
g^{kb}\bigr)\bigr]_{,b}.
\end{equation}
$_E t_i^{~k}$ are components of the canonical energy-momentum
pseudotensor for gravitational field, and $_F {U_i^{~[kl]}}=(-) _F {U_i^{~[lk]}}$ are
Freud's superpotential.
\begin{equation}
_E K_i^{~k}: = \sqrt{\vert g\vert}\bigl(T_i^{~k} + _E
t_i^{~k}\bigr)
\end{equation}
are components of the {\it Einstein canonical energy-momentum complex
for matter and gravity} in {\bf GR}.

In consequence of (1) the complex $_E K_i^{~k}$satisfies the {\it continuity equations} (=local
conservation laws)
\begin{equation}
{_E K_i^{~k}}_{,k} =_F {U_i^{~[kl]}}_{,lk} = 0.
\end{equation}

In very special cases, e.g., in the case of a closed system, {\bf CS}, \cite{Moel} one can
obtain from these continuity equations the four reasonable integral conservation laws
energy and momentum.

From (5) one can also obtain, by successive differentiation,
infinitely many other continuity equations
\begin{equation}
[\sqrt{\vert g\vert}\bigl(T_i^{~k} + _E
t_i^{~k}\bigr)]_{,klm...rst} = _F {U_i^{~[kl]}}_{lkm...rst}  = 0,
\end{equation}
especially, one can get from (5) the following continuity
equations
\begin{equation}
[\delta^{ab}\sqrt{\vert g\vert}\bigl(T_i^{~k} + _E
t_i^{~k}\bigr)]_{,kab} = _F {U_i^{~[kl]}}_{,lkab}{}\delta^{ab} =0,
\end{equation}
or, by using Schwarz's Lemma, the equations
\begin{equation}
\bigl\{[\delta^{ab}\sqrt{\vert g\vert}\bigl(T_i^{~k} + _E
t_i^{~k}\bigr)]_{,ab}\bigr\}_{,k} = [\bigl(\delta^{ab}~ _F
{U_i^{[kl]}}\bigr)_{,lab}]_{,k} = 0.
\end{equation}

After introducing differential operator
\begin{equation}
\Delta^{(4)}:= \partial_o^{~2} +\partial_1^{~2} + \partial_2^{~2} +
\partial_3^{~2} \equiv \delta^{ab} \partial_a \partial_b
\end{equation}
one can write the equations (8) in the form
\begin{equation}
\bigl(\Delta^{(4)}~ _E K_i^{~k}\bigr)_{,k} =
\bigl(\Delta^{(4)}~_F{U_i^{~[kl]}}_{,l}\bigr)_{,k} = 0.
\end{equation}

The continuity equations (6), which follow from (5), {\it have, in
general, no new physical meaning} except the case (7)-(8) in which
 analytic 4-Laplacian $\Delta^{(4)} =\delta^{ab} \partial_a \partial_b$
 occurs.

Namely, from the constructive definition of the canonical
superenergy tensors, $_g S_i^{~k},~~ _m S_i^{~k}$, gravity and matter,
given in \cite{Gar1, Gar2} (See also Appendix), one can easily see
that these tensors are exactly 4-dimensional analytic Laplacians
\begin{equation}
\Delta^{(4)} T_i^{~k}(P)= \delta^{ab}
{T_i^{~k}}_{,ab}(P),~~\Delta^{(4)}~ _E t_i^{~k}(P) = \delta^{ab}{_E
t_i^{~k}}_{,ab}(P),
\end{equation}
calculated in Riemann normal coordinates, {\bf RNC(P)}, taken at
the point {\bf P} [{\bf P} = beginning  of the {\bf RNC(P)}], and
then, expressed covariantly in terms of normal tensors, tensor
extensions \cite{Sch, Veb}, and 4-velocity of a fiducial observer
{\bf O}, which is at rest at point {\bf P}.

Therefore, the quantity on the left hand side of the continuity
equations (8), i.e.,
\begin{equation}
\Delta^{(4)} _E K_i^{~k} = \bigl[ \delta^{ab}\sqrt{\vert
g\vert}\bigl(T_i^{~k} + _E t_i^{~k}\bigr)\bigr]_{,ab},
\end{equation}
when taken at beginning {\bf P} of the {\bf RNC(P)} and
covariantly expressed, gives us exactly the total canonical
superenergy tensor, matter and gravity, $_m S_i^{~k}(P;v^l) + _g S_i^{~k}(P;v^l)$
(See \cite{Gar1, Gar2} and Appendix).

From these reasons we call the analytic Laplacian
\begin{equation}
\bigl(\delta^{ab}~ _E K_i^{~k}\bigr)_{,ab} = \bigl(\delta^{ab}
{_F U_i^{~[kl]}}_{,l}\bigr)_{,ab}
\end{equation}
{\it the canonical superenergy complex}, matter and gravitation,
and denote it $_K S_i^{~k}$.\footnote{Complex because the
components of $_K S_i^{~k}$, like the components $_E K_i^{~k}$, do not form any tensor.}

From (8) we see that the complex, $_K S_i^{~k}$, like the canonical
energy-momentum complex, $_E K_i^{~k}$, satisfies continuity
equations
\begin{equation}
_K S_i^{~k}{,k} = 0.
\end{equation}

The components of the complex, $_K S_i^{~k}$, have the same
dimensions as components of the canonical superenergy tensors,
i.e., they have dimensions  of the energy-momentum divided
by $m^2$: $[_K S_i^{~k}] = {[T_i^{~k}]\over m^2}$.

We would like to emphasize that the canonical superenergy density,
$\epsilon_s$, calculated from superenergy complex or from superenergy tensors
exactly corresponds to Appel's energy of acceleration which plays
important role in classical mechanics (See, e.g., \cite{Gar1,
Gar2, Appel}).

Consequently, the canonical superenergy complex, $_K S_i^{~k}$, and
the continuity equations (14) surely can have a physical meaning
(Like the canonical superenergy tensors, $_g S_i^{~k}, ~~_m
S_i^{~k}$, gravitation and matter).

In analogical way one can introduce into {\bf GR} {\it the
canonical angular supermomentum complex}, $_K S^{ijk} = (-) _K
S^{jik}$, matter and gravitation.

Namely, we start from the canonical angular momentum complex, $M^{ijk} = (-)
M^{jik}$, in {\bf GR} \cite{Ber, Gar3}
\begin{eqnarray}
M^{ijk} &=& (-) M^{jik} := x^i _{BT} K^{jk} - x^j _{BT} K^{ik} +
_F U^{i[jk]}- _F U^{j[ik]}\nonumber\\
&=&  \bigl(x^i _F U^{j[kl]} - x^j _F U^{i[kl]}\bigr)_{,l}  =:
{M^{[ij][kl]}}_{,l}.
\end{eqnarray}

Here
\begin{equation}
_{BT}  K^{ik} := g^{ij}~ _E K_j^{~k} + _F
U_j^{~[kl]}{}g^{ij}_{~~,l}=\sqrt{vert g\vert}\bigl(T^{ik} + _{BT} t^{ik}\bigr) \not= _{BT} K^{ki}
\end{equation}
are components of the {\it Bergmann-Thomson energy-momentum
complex in} {\bf GR} \cite{Ber, Gar3}, $_F U^{i[jk]}:= g^{il}~
_FU_l^{~[jk]}$, and double antysymmetric quantity
\begin{equation}
M^{[ij][kl]}:= x^i _F U^{j[kl]} - x^j _F U^{i[kl]}
\end{equation}
is {\it angular momentum superpotential}.

$_{BT} t^{ik}\not= _{BT} t^{ki}$ is the {\it Bergmann-Thomson}
gravitational energy- momentum pseudotensor (See Appendix).

The non-tensorial (that's why ``complex'') complex, $M^{ijk} =(-)
M^{jik}$, satisfies, in consequence of (15), the continuity equations
\begin{equation}
M^{ijk}_{~~~,k} = M^{[ij][kl]}_{~~~~~~~~,lk} = 0.
\end{equation}

From (18) one can get for a closed system the reasonable six
integral conservation laws for angular momentum \cite{Moel, Ber,
Gar3}.

We introduce the {\it canonical angular supermomentum complex} in
{\bf GR} (also non-tensorial), $_K S^{ijk} =(-) _K S^{jik}$, by
definition
\begin{equation}
_K S^{ijk} := \bigl(\delta^{ab} M^{ijk}\bigr)_{,ab} = \Delta^{(4)} M^{ijk}.
\end{equation}

As consequence of the continuity equations (18) and Schwarz's
Lemma one has
\begin{equation}
_K S^{ijk}_{~~~,k} =0,
\end{equation}
i.e., one has continuity equations for canonical angular
supermomentum.

The quantity (19), when taken at the beginning {\bf P} of the
{\bf RNC(P)} and expressed covariantly in terms of the normal
tensors, tensors extensions, and 4-velocity of a fiducial observer
{\bf O} at rest in {\bf P} exactly gives our total canonical
angular supermomentum tensor
\begin{equation}
S^{ikl}=(-) S^{kil} = _g S^{ikl} + _m S^{ikl},
\end{equation}
gravitation and matter \cite{Gar1, Gar2} (See also Appendix).

Thus, the name {\it angular supermomentum complex} for quantity
(19) is justified.

In the following we will apply the above introduced canonical
superenergy  and angular supermomentum complexes to analyze
a closed gravitational system, {\bf CS}, to analyze Trautman's
radiative spacetimes, {\bf TRS}, and to analyze Friedman
universes, {\bf FU}. We will compare the obtained results with our
results obtained in past by performing analogical
analyzis with help of the canonical superenergy tensors and the
canonical angular supermomentum tensors.

We will see that the canonical superenergy and angular
supermomentum complexes are, in some sense, complementary quantities to the
canonical superenergy and angular momentum tensors. Namely, these complexes enable us
global analysis of the solutions to the Einstein equations. But
this analysis is coordinate--dependent. On the other hand, the
canonical superenergy and angular supermomentum tensors are
suitable to coordinate-independent, local analysis of such
solutions.
\section{Application to a closed gravitational system}

Henceforth we will use {\it geometrical units} in which $G=c = 1$.

By {\it closed system}, {\bf CS}, we mean an isolated material
system which admits global coordinates $(t,x,y,z)$ in which metric
components, $g_{ik}$, have the form

\begin{eqnarray}
g_{ik} & = &     \eta_{ik} + h_{ik},\nonumber\\
h_{ik} &=& 0({1\over r}), ~~g_{ik,l} = 0({1\over r^2}),
\end{eqnarray}
where
\begin{equation}
r^2 = x^2 + y^2 + z^2,
\end{equation}
and $\eta_{ik} = diag(1,-1,-1,-1)$.

The coordinates $(t,x,y,z)$ are called {\it asymptotically flat
Bondi-Sachs coordinates}. (For more detailed description of behaviour
of the metric components for a {\bf CS}, see \cite{Moel}).

Integrating the continuity equations (14) over the spatial section $x^0 = t =
const$, one gets (with help of the Stokes integral theorem, see,
e.g.,  \cite{LL, Moel})
\begin{equation}
{d\over dt}\oint\limits_{\partial x^0}{\bigl\{\delta^{ab} {_F
U_i^{~[0\alpha]}}_{ ,ab}{} n_{\alpha} d^2\sigma\bigr\}}=(-)
\oint\limits_{\partial x^0}{ \bigl\{\delta^{ab} {_F U_i^{~[\alpha
l]}}_{,lab}{} n_{\alpha} d^2\sigma\bigr\}}.
\end{equation}
Here, $\partial x^0$, means 2-dimensional boundary of the spatial
slice $x^0 = t = const$; $n_{\alpha}$ denotes spatial components of
the unit normal $\vec n$ to this boundary directed outside,
and $d^2\sigma = r^2\sin\theta d\theta d\varphi$.

The equations (24) represent the four integral conservation laws
for canonical superenergy $_K S_0$ and supermomentum $_K
S_{\alpha}$:
\begin{equation}
{d\over dt}{_K S_i} = (-) \oint\limits_{\partial x^0}{{_K
S_i^{~\alpha}}n_{\alpha} d^2\sigma},
\end{equation}
where
\begin{equation}
_K S_i := \int\limits_{x^0 = const}{\bigl(\delta^{ab}{_F
U_i^{~[0\alpha]}}_{,ab}\bigr)_{,\alpha}d^3v} =
\oint\limits_{\partial x^0}{\delta^{ab}{_F U_i^{~[0\alpha]}}_{,ab}
n_{\alpha} d^2\sigma}
\end{equation}
and
\begin{equation}
_K S_i^{~\alpha} := \delta^{ab} {_F U_i^{~[\alpha l]}}_{,lab}.
\end{equation}

One can easily calculate (See, e.g., \cite{Moel}) that for a
closed system {\bf CS}
\begin{equation}
{_F U_i^{~[0\alpha]}}_{,ab}  n_{\alpha} = O({1\over r^4}),
\end{equation}
\begin{equation}
{_F U_i^{~[\alpha l]}}_{,lab} = O({1\over r^4}).
\end{equation}
Therefore, the {\it integral conservation laws} (24) [or (25)]
{\it trivialize} to the form $0=0$ in the case because $d^2\sigma =
O(r^2)$, and  {\it have no physical meaning}.

This is reasonable result because we have in
the case four ordinary integral conservation laws of the global
energy-momentum (See, e.g., \cite{Moel}), and, in consequence , we
needn't any additional integral conservation laws which would have been
physically valid.

In a similar way the continuity equations (20) lead us finally to
the following six integral equalities
\begin{eqnarray}
{d\over dt}\int\limits_{x^0 =const}{{_K S^{ij0}}d^3v} &=& (-)
\int\limits_{x^0 = const}{{_K S^{ij\alpha}}_{,\alpha}
d^3v}\nonumber\\
&=& (-) \oint\limits_{\partial x^0} {{_K  S^{ij\alpha}} n_{\alpha}
d^2\sigma},
\end{eqnarray}

or
\begin{equation}
{d\over dt}M^{ij} =(-) \oint\limits_{\partial x^0}{{_K
S^{ij\alpha}}n_{\alpha}d^2\sigma}.
\end{equation}

In extended form
\begin{eqnarray}
M^{ij} &=& (-) M^{ji} :=  \int\limits_{x^0 = const}{{_K
S^{ij0}}d^3v} = \int\limits_{x^0 = const}{\bigl(\delta^{ab}
M^{ij0}\bigr)_{,ab} d^3v}\nonumber\\
&=& \oint\limits_{\partial x^0}{\bigl(\delta^{ab}
{M^{[ij][0\alpha]}}_{,ab}\bigr) n_{\alpha}d^2\sigma},
\end{eqnarray}
and
\begin{equation}
_K S^{ij\alpha} := \delta^{ab} {M^{[ij][\alpha l]}}_{,lab}.
\end{equation}

The equations (31) could give us six integral conservation laws of
the canonical angular supermomentum $M^{ij} = (-) M^{ji}$. But,
one can calculate that for a {\bf CS} (See, eg., \cite{Moel})
\begin{equation}
{_F U^{j[0\alpha]}}_{,b} n_{\alpha} = O(r^{-3}), ~~{_F
U^{j[0\alpha]}}_{,ab} n_{\alpha} = O(r^{-4}).
\end{equation}
Substituting these asymptotics into integrals in (31) we obtain
\begin{equation}
M^{ij} = (-) M^{ji} = 0, ~~\oint\limits_{\partial
x^0}{\delta^{ab}{M^{[ij][\alpha l]}}_{,lab} n_{\alpha} d^2\sigma} =
0,
\end{equation}
i.e., we again obtain six trivial equalities $0=0$.

Like as it was in the case of the canonical superenergy and supermomentum
the last result is very satisfactory because we have here already
six integral conservation laws for angular momentum and
we needn't any other nontrivial integral conserwation laws.
\section{Application to Trautman's radiative spacetimes}

By Trautman's radiative spacetime, {\bf TRS}, we mean vacuum
solution to the Einstein equations admitting asymptotically flat
coordinates $(t,x,y,z)$ in which one has
\begin{eqnarray}
g_{ik} &=& \eta_{ik} + O(r^{-1}), ~~g_{ik,l} = I_{ik} k_l +
O(r^{-2}),~~I_{ik} = O(r^{-1}),\nonumber\\
\bigl(I_{ik} &-& {1\over 2}\eta_{ik}{}\eta^{ab}{} I_{ab}\bigr)k^k
= O(r^{-2}),\nonumber\\
g_{ik,lm} &=& J_{ik}k_l k_m + O(r^{-2}), ~~J_{ik} = J_{ki} =
O(r^{-1}),\nonumber\\
\Gamma ^a_{~bc}& = & \Gamma^a_{~cb} = O(r^{-1}), ~~R_{iklm} =
O(r^{-1}).
\end{eqnarray}
$k^i$ are components of a null vector directed to scri-plus
,$S^+$, and $r = \sqrt{x^2 + y^2 + z^2}$ (For more details, see
e.g., \cite{Moel,Traut,And}).

Trautman's radiative spacetimes admit outgoing gravitational
radiation.

One can calculate (See, e.g.,\cite{Moel}) that for, {\bf TRS}, one
has in the coordinates $(t,x,y,z)$
\begin{eqnarray}
{_F U_i^{~[kl]}}_{,l}& =& O(r^{-2}), ~~~{_F
U_i^{~[0\alpha]}}n_{\alpha} = O(r^{-2}),\nonumber\\
{_F U_i^{~[0\alpha]}}_{,\alpha}&=& O(r^{-2}).
\end{eqnarray}

From this it follows that
\begin{eqnarray}
\bigl[\bigl(\delta^{ab}{_F
U_i^{~[o\alpha]}}\bigr)_{,ab}\bigr]_{,\alpha} &=& O(r^{-2}),
~~\bigl[\bigl(\delta^{ab} {_F U_i^{~[0\alpha]}}\bigr)_{, ab}\bigr]
n_{\alpha}= O(r^{-2}),\nonumber\\
{_F U_i^{~[0\alpha]}}_{,b\alpha}&=& O(r^{-2}).
\end{eqnarray}
 Using the above asymptotics one obtains that the integrals (26)
 on $S_i$ are convergent  and different from zero but they depend
 on time.

The integrals on the right hand of (24) (or (25)) are also convergent.

So, in the case of {\bf TRS} the equations (24) (or (25))do not trivialize
to the form $0=0$ but they give us the laws of the temporal change
of the integral canonical superenergetic quantities $S_i$.

Similar situation we have in the case for the canonical energetic
quantities (See, e.g., \cite{Moel,And}).

Therefore, in {\bf TRS}, where we have convergent but depended on
time integral energetic quantities, like energy and momentum, the
integral superenergetic quantities become nontrivial and can be
physically valid.

Concerning components of the angular supermomentum complex $_K S^{ijk} = (-) _K
S^{jik}$, one can easily see from the extended formulas
(30)-(33) and from  the asymptotics (38) that the integrals on $M^{ij} = (-) M^{ji}$
{\it are divergent} in {\bf TRS} [$\bigl(\delta^{ab}{M^{[ij][0\alpha]}}_{,ab}\bigr)n_{\alpha}
 = O(r^{-1})$ in the case].

Thus, we have in the case the same situation as in the case of the
ordinary angular momentum.
\section{Formal application of the canonical superenergy complex $_K S_i^{~k}$
to analyze Friedman universes}

Here we confine to canonical superenergy complex only.

If one formally uses the canonical superenergy complex
\begin{equation}
_K S_i^{~k} = \bigl(\delta^{ab}{}_K K_i^{~k}\bigr)_{,ab}= {_F
U_i^{~[kl]}}_{,lab}\delta^{ab}
\end{equation}
to analyze Friedman universess in ``Carthesian'' coordinates $(t,x,y,z)$
in which Friedman line element reads
\begin{equation}
ds^2 = dt^2 - {R^2(t)\bigl(dx^2 + dy^2 + dz^2\bigr)\over\bigl[1 +
{k\over 4}\bigl(x^2 + y^2 + z^2\bigr)\bigr]^2},
\end{equation}
where the {\it curvature index} $k = 0,\pm 1$, and $R = R(t)$ is
the so-called {\it scale factor}, then, after some calculations,
one can see that this complex is better to this aim than the
canonical energy-momentum complex $_E K_i^{~k} = {_F
U_i^{~[kl]}}_{,l}$ e.g., it better suits to singularity analysis in Friedman universes
than the complex $_E K_i^{~k}$ \cite{Gar1,Gar2}. (See also below).

In the flat case $k =0$ all ``densities'' of the canonical
superenergetic quantities {\it trivially vanish} being multiplied
by $k=0$. In consequence, the formally calculated integral
canonical superenergy quantities also trivially vanish in the
case.

It is interesting that the same result we have for canonical
energy-momentum for flat Friedman universes (See, e.g.,
\cite{Gar1}).

In the cases $k=\pm 1$ the ``densities'' of the canonical
superenergetic quantities are different from zero and all go to
(-)infinity when $R(t) \longrightarrow 0^{+}$.

Therefore, these ``densities'' can be used to analysis of the
initial singularities in these cases.

We would like to remark that the canonical energetic quantities
are not relevant to this aim because all their ``densities'' go
to zero when the scale factor goes to zero (See, e.g.,
\cite{Gar1}).

The formally calculated integral canonical superenergetic
quantities for Friedman universes having $k=\pm 1$ read
\begin{equation}
S_0 = \int\limits_{t = const}{_K S_0^{~0}  d^3v}
=\cases{{49\chi\pi^2 R\over 8}>0,& if k=1;\cr (-)\infty,& if k =(-)1\cr},
\end{equation}
and
\begin{equation}
S_{\beta}:= \int\limits_{t = const}{_K S_{\beta}^{~0} dxdydz} = 0.
\end{equation}
Here $S_{\beta}$ are components of integral linear supermomentum
and $\chi = {1\over 16\pi}$.

The above results are very similar to the results obtained in past
for integral energetic quantities  (except $E = P_0 = 0 ~~if~~
k=1$) \cite{Gar1}.

We must emphasize that there exist natural objections against
integral quantities for Friedman universes because these universes
are not asymptotically flat: they are conformally flat only. In
consequence, the integral quantities of the Friedman universes
{\it are not measurable}. Therefore, considerations of these
quantities for Friedman universes can have some mathematical meaning only.

Finishing this Section we would like to remark that from both,
geometrical and physical points of view, the using of the our
canonical superenergy tensors $_g S_i^{~k}, ~~_m S_i^{~k}$,
gravitation and matter, to analyze Friedman universe is much more
reasonable than the using to this goal the canonical superenergy
complex. For example, the canonical superenergy densities,
$\epsilon_s$, are positive definite scalars for all {\bf FU} and
they are singular when $R(t)\longrightarrow 0^+$.
\section{Canonical superenergy and angular supermomentum complexes
versus canonical superenergy and angular supermomentum tensors}

In past we have introduced the canonical superenergy and angular
supermomentum tensors, matter and gravitation and total, by using
special averaging of the differences of the canonical
energy-momentum in Riemann normal coordinates, {\bf
RNC(P)},\cite{Gar1} (See also Appendix). {\bf P} is the beginning of these coordinates.

These tensors, constructed pointwise, were very suitable to local,
coordinate independent analysis of the gravitational field, and
also to analyze matter field. Moreover, the canonical superenergy
tensor for gravitational field, $_g S_i^{~k}(P;v^l)$, gave us some
substitute of the non-existing gravitational energy-momentum
tensor.

The constructive definitions of the canonical superenergy tensors,
$_g S_i^{~k}, ~~_g S^{ikl}, ~~_m S_i^{~k}, ~~_m S^{ikl}$,
immediately led us at first to 4-dimensional ``Laplacians'' $\delta^{ab}\partial_a \partial_b$
at point {\bf P} of the averaged fields. Then, after expressing these
``Laplacians'' covariantly in terms of curvature tensor and its
comitants, in terms of 4-velocity of a fiducial observer {\bf O}
which is at rest at point {\bf P} and in terms of covariant
derivatives of matter tensor, we have obtained the our superenergy
and angular supermomentum tensors $_g S_i^{~k}, ~~_m S_i^{~k}, ~~ _g S^{ikl}, ~~_m
S^{ikl}$.

About four years ago we have observed that these tensors can be also
obtained more easily from the canonical superenergy and angular
supermomentum complexes $_K S_i^{~k}, ~~_K S^{ikl} = (-) _K
S^{kil}$.

With this aim it is sufficient to take the components of these
complexes at point {\bf P}= beginning of the {\bf RNC(P)} and express
them covariantly by using special properties of the {\bf
RNC(P)}\cite{Sch, Veb}.

In this paper we have defined the canonical superenergy and
angular supermomentum complexes $_K S_i^{~k}, ~~~_K S^{ikl} = (-) _K
S^{kil}$, and applied them to analyze of a {\bf CS}, {\bf TRS},
and to analyze (superenergetic only) Friedman universes, {\bf FU}.

In the case of a  {\bf CS} the integral canonical superenergetic
quantities satisfy four trivial conservation laws $0=0$, and in
the case  {\bf TRS} they behave very similar to the integral
canonical energetic quantities.

On the other hand, it seems that the canonical superenergy complex
$_K S_i^{~k}$ gives better tool to analysis Friedman universes
in ``Carthesian'' coordinates $(t,x,y,z)$ than the canonical
energy-momentum complex $_E K_i^{~k}$ \footnote{But much worse tool
than the canonical superenergy tensors $_g S_i^{~k}, ~~~_m
S_i^{~k}$}.

In our previous papers \cite{Gar1, Gar2} we have applied the
canonical superenergy tensors, matter and gravitation, $_m S_i ^{~k}, ~~~_g S_i
^{~k}$, to local, and in some special cases,  also  to global
analysis of the very known solutions to the Einstein equations.

The obtained results were interesting \cite {Gar1, Gar2} (See also
Appendix). In general one can say that the canonical superenergy and angular
supermomentu tensors give better tool to local analysis of these
solutions. Moreover, the canonical angular supermomentum tensors
,as being independent of any radius vector, lead to better
convergence of the suitable integrals in  {\bf TRS}.

Recently we have proposed \cite{Gar1} to use the total canonical
superenergy density $\epsilon_s := \bigl(_g S_i^{~k} + _m S_i
^{~k}\bigr)v^i{}v_k$ to study gravitational stability of the
solutions to the Einstein equations.

Comparing the two approaches to superenergy:
\begin{enumerate}
\item Canonical superenergy tensors
$_g S_i^{~k}, ~~_m S_i^{~k}; ~~S_i^{~k}:= _g S_i^{~k} + _m
S_i^{~k}$,
\item Canonical superenergy complex $_K S _i^{~k}$, gravity and
matter,
\end{enumerate}
one can conclude that the canonical superenrgy tensors give better
tool to local (but in some cases also to global) analysis
of the solutions to the Einstein equations than the canonical
superenergy complex. It is mainly because they are tensors.

The canonical superenergy tensors, as a result of some averaging,
do not satisfy in general any conserwation laws, local or global.
It is a defect of these tensors. But they refer to real
gravitational field only and it is a positive property.

One can look on these tensors as on some kind of quasilocal
quantities which are not conserved.

Contrary, the canonical superenergy complex $_K S_i^{~k}$
satisfies conservation laws but it is coordinate
dependent, non-tensorial quantity and, therefore, it can be
reasonably use in special coordinates only, e.g.,to global analysis of
an asymptotically flat spacetime in an asymptotically flat
coordinates.

The analogical properties possesses the canonical energy-momentum
complex $_E K_i^{~k}$.

One can see on the canonical superenergy complex $_K S_i^{~k}$ as
on a conserved quasilocal quantity.

The complex $_K S_i^{~k}$, like as the canonical energy-momentum
complex $_E K_i^{~k}$, is a {\it mixture of real and fictive
(=inertial)} gravitational fields.

Concerning canonical angular supermomentum complex $_K S^{ikl} = (-) _K
S^{kil}$ one can say that it satisfies continuity equations (30)-(33),
and, in the case of a {\bf CS} it leads to six, trivial, integral
conservation laws $0=0$. In {\bf TRS} this complex behaves very
alike to the canonical angular momentu complex $M^{ijk} = (-)
M^{jik}$ leading to divergent integrals.

The canonical angular supermomentum tensors gravitation and
matter, $_g S^{ikl}, ~~_m S^{ikl}$, and total $S^{ikl} := _g S^{ikl} + _m
S^{ikl}$, introduced in our previous papers \cite{Gar1, Gar2} (See
also Appendix) give better tool to analyze gravitational and
matter fields than the canonical angular supermomentum complex $_K S^{ikl} = (-) _K
S^{kil}$ despite that they do not satisfy any continuity
equations. For example, they are tensors  and lead to better
convergence integrals in {\bf TRS} owing the fact that they {\it
do not depend on components of any radius-vector} (See, e.g.,
\cite{Gar1, Gar2} and Appendix).

Similarly as it was in the case canonical superenergy, one can
interpret the complex $_K S ^{ikl}$  as a conserved quasilocal
quantity, and the tensors $_g S^{ikl}, ~~_m S ^{ikl}, ~~S^{ikl}$
as  non-conserved quasilocal quantities.

One can consider differences
\begin{eqnarray}
R_i^{~K} &:=& _K S_i^{~k} - S_i^{~k},\nonumber\cr
R^{ikl}& :=& _K S^{ikl} - S^{ikl},
\end{eqnarray}
and connect them with {\it inertial forces}, i.e., with fictive
gravitational field.

Here
\begin{equation}
S_i^{~k} := _g S_i^{~k} + _m S_i^{~k}, ~~S^{ikl} := _g S ^{ikl} +
_m S ^{ikl}
\end{equation}
mean components of the total superenergy and total angular
supermomentum tensors matter and gravitation respectively.

\section{Appendix}

Here we remind  constructive definition of the superenergy tensor $S_a^{~b}$ and analogical
definition of the angular supermomentum tensor $S^{ikl}$  applicable to
gravitational field and to any matter field. The definitions use
{\it locally Minkowskian structure} of the spacetime
and, therefore, they  fail in a spacetime with torsion, e.g., in Riemann-Cartan
spacetime.

Let us start with the superenergy tensor.

In the normal Riemann coordinates ,{\bf NRC(P)}, we define
(pointwise)
\begin{equation}
S_{(a)}^{~~~(b)}(P) = S_a^{~b} :=(-) \displaystyle\lim_{\Omega\to
P}{\int\limits_{\Omega}\biggl[T_{(a)}^{~~~(b)}(y) - T_{(a)}^{
~~~(b)} (P)\biggr]d\Omega\over 1/2\int\limits_{\Omega}\sigma(P;y)
d\Omega},
\end{equation}
where
\begin{eqnarray}
T_{(a)}^{~~~(b)}(y) &:=& T_i^{~k}(y)e^i_{~(a)}(y)
e_k^{~(b)}(y),\nonumber\\
T_{(a)}^{~~~(b)}(P)&:=& T_i^{~k}(P) e^i_{~(a)}(P)e_k^{~(b)}(P) =
T_a^{~b}(P)
\end{eqnarray}
are {\it physical or tetrad components} of the pseudotensor or
tensor field which describes an energy-momentum distribution, and $\bigl\{y^i\bigr\}$
are normal coordinates. $e^i_{~(a)}(y), e_k^{~(b)} (y)$ denote an
orthonormal tetrad $e^i_{~(a)}(P) = \delta_a^i$ and its dual $e_k^{~(a)}(P) = \delta_k^a
$, paralelly propagated along geodesics through {\bf P} ({\bf P} is the origin
of the {\bf NRC(P)}).

We have
\begin{equation}
e^i_{~(a)}(y) e_i^{~(b)}(y) = \delta_a^b.
\end{equation}

For a sufficiently small 4-dimensional domain $\Omega$ which
surrounds {\bf P} we require
\begin{equation}
\int\limits_{\Omega}{y^i d\Omega} = 0, ~~\int\limits_{\Omega}{y^i
y^k d\Omega} = \delta^{ik} M,
\end{equation}
where
\begin{equation}
M = \int\limits_{\Omega}{(y^0)^2 d\Omega} =
\int\limits_{\Omega}{(y^1)^2 d\Omega} =
\int\limits_{\Omega}{(y^2)^2
d\Omega}=\int\limits_{\Omega}{(y^3)^2 d\Omega},
\end{equation}
is a common value of the moments of inertia of the domain $\Omega$
with respect to the subspaces $y^i = 0,~~(i= 0,1,2,3)$.

As $\Omega$ we can take, e.g., a  sufficiently small analytic ball centered
at {\bf P}:
\begin{equation}
(y^0)^2 + (y^1)^2 + (y^2)^2 + (y^3)^2 \leq R^2,
\end{equation}
which for an auxiliary positive-definite metric
\begin{equation}
h^{ik} := 2 v^i v^k - g^{ik},
\end{equation}
can be written in the form
\begin{equation}
h_{ik}y^i y^k \leq R^2.
\end{equation}
A fiducial observer {\bf O} is at rest at the beginning {\bf P}
of the Riemann normal coordinates ,{\bf NRC(P)}, and its four-
velocity is $v^i =\ast~ \delta^i_o.$ $=\ast$ means that
equation is valid only in special coordinates.
$\sigma(P;y)$ denotes the two-point {\it world function}
introduced by J.L. Synge \cite{Synge}:
\begin{equation}
\sigma(P;y) =\ast {1\over 2}\bigl(y^{o^2} - y^{1^2} - y^{2^2}
-y^{3^2}\bigr).
\end{equation}

The world function $\sigma(P;y)$ can be defined covariantly by the
{\it eikonal-like equation} \cite{Synge}
\begin{equation}
g^{ik} \sigma_{,i} \sigma_{,k} = 2\sigma,
~~\sigma_{,i} := \partial_i\sigma,
\end{equation}
together with requirements
\begin{equation}
\sigma(P;P) = 0, ~~\partial_i\sigma(P;P) = 0.
\end{equation}

The ball $\Omega$ can also be given by the inequality
\begin{equation}
h^{ik}\sigma_{,i} \sigma_{,k} \leq R^2.
\end{equation}

Tetrad components and normal components are equal at {\bf P}, so,
we will write the components of any quantity attached to {\bf P}
without tetrad brackets, e.g., we will write $S_a^{~b}(P)$
instead of $S_{(a)}^{~~~(b)}(P)$ and so on.

If $T_i^{~k}(y)$ are the components of an energy-momentum tensor
of matter, then we get from (44)
\begin{equation}
_m S_a^{~b}(P;v^l) = \bigl(2{\hat v}^l {\hat v}^m - {\hat g}^{lm}\bigr) \nabla_l \nabla_m {}
{\hat T}_a^{~b} = {\hat h}^{lm}\nabla_l \nabla_m {}{\hat T}_a^{~b}.
\end{equation}
Hat over a quantity denotes its value at {\bf P}, and $\nabla$
means covariant derivative.

Tensor $_m S_a^{~b}(P;v^l)$ is called {\it the canonical superenergy tensor for matter}.

For the gravitational field, substitution of the canonical
Einstein energy-momentum pseudotensor as $T_i^{~k}$ into (44) gives
\begin{equation}
_g S_a^{~b}(P;v^l) = {\hat h}^{lm} {\hat W}_a^{~b}{}_{lm},
\end{equation}
where
\begin{eqnarray}
{W_a^{~b}}{}_{lm}&=& {2\alpha\over 9}\bigl[B^b_{~alm} +
P^b_{~alm}\nonumber\\
&-& {1\over 2}\delta^b_a R^{ijk}_{~~~m}\bigl(R_{ijkl} +
R_{ikjl}\bigr) + 2\delta_a^b{\beta}^2 E_{(l\vert g}{}E^g_{~\vert
m)}\nonumber\\
&-& 3 {\beta}^2 E_{a(l\vert}{}E^b_{~\vert m)} + 2\beta
R^b_{~(a\vert g\vert l)}{}E^g_{~m}\bigr].
\end{eqnarray}
Here $\alpha = {c^4\over 16\pi G} = {1\over 2\beta}$
\footnote{In geometrized units $\alpha = {1\over 16\pi}$},
and
\begin{equation}
E_i^{~k} := T_i^{~k} - {1\over 2}\delta_i^k T
\end{equation}
is the {\it modified energy-momentum tensor} of matter \footnote{In
terms of $E_i^{~k}$ Einstein equations read $R_i^{~k} = \beta
E_i^{~k}$.}.

On the other hand
\begin{equation}
B^b_{~alm} := 2R^{bik}_{~~~(l\vert}{}R_{aik\vert m)}-{1\over
2}\delta_a^b{} R^{ijk}_{~~~l}{}R_{ijkm}
\end{equation}
are components of the {\it Bel-Robinson tensor} ({\bf BRT}),
while
\begin{equation}
P^b_{~alm}:= 2R^{bik}_{~~~(l\vert}{}R_{aki\vert m)}-{1\over
2} \delta_a^b{}R^{jik}_{~~~l}{}R_{jkim}
\end{equation}
is the Bel-Robinson tensor with  ``transposed'' indices $(ik)$.

In vacuum $_g S_a^{~b}(P;v^l)$ takes the simpler form
\begin{equation}
_g S_a^{~b}(P;v^l) = {8\alpha\over 9} {\hat h}^{lm}\bigl({\hat
C}^{bik}_{~~~(l\vert}{}{\hat C}_{aik\vert m)} -{1\over
2}\delta_a^b {\hat C}^{i(kp)}_{~~~~~(l\vert}{}{\hat C}_{ikp\vert
m)}\bigr).
\end{equation}
Here $C^a_{~blm}$ denote components of the {\it Weyl tensor}.

The canonical angular supermomentum tensors, analogous to the case
of the canonical superenergy tensors, can be considered as {\it
some substitutes} for the angular momentum tensors of matter
and gravitation which do not exist in {\bf GR}.

The constructive definition of these tensors is the following.
In analogy to the definition (44) of the canonical superenergy
tensors we define  in {\bf RNC(P)}
\begin{equation}
S^{(a)(b)(c)} = S^{abc}(P) := -\displaystyle\lim_{\Omega\to
P}{\int\limits_{\Omega}\bigl[M^{(a)(b)(c)}(y) - M^{(a)(b)(c)}(P)
\bigr]d\Omega\over 1/2\int\limits_{\Omega}\sigma(P;y)d\Omega},
\end{equation}
where
\begin{equation}
M^{(a)(b)(c)}(y) := M^{ikl}(y)
e_i^{~(a)}(y){}e_k^{~(b)}(y){}e_l^{~(c)}(y),
\end{equation}
\begin{equation}
M^{(a)(b)(c)}(P):= M^{ikl}(P)
e_i^{~(a)}(P){}e_k^{~(b)}(P){}e_l^{~(c)}(P) =
M^{ikl}\delta_i^a{}\delta_k^b{}\delta_l^c = M^{abc}(P),
\end{equation}
are the physical (or tetrad) components of the field  $M^{ikl}(y) = - M^{kil}(y)$
which describes angular momentum densities.

$e^i_{~(a)}(y), ~~~e_k^{~(b)}(y)$ mean orthonormal base such that
$e^i_{~(a)}(P) = \delta_a^b$ and its dual {\it parallelly propagated
along geodesics through} {\bf P}. $\Omega$ is an already defined  sufficiently small
analytic ball with centre {\bf P}.

At the point {\bf P}we have equality of the tetrad and normal
components. We use this fact and omit tetrad brackets for indices
of any quantity attached at the point {\bf P}; for example, we
write $S^{abc}(P)$ instead of $S^{(a)(b)(c)}(P)$ and so on.

For matter, as $_m M^{ikl}(y) = (-) _m M^{kil}(y)$ we take
\begin{equation}
_m M^{ikl}(y) = \sqrt{\vert g\vert}\bigl(y^i T^{kl} - y^k
T^{il}\bigr),
\end{equation}
where $T^{ik} = T^{ki}$ are the components of a symmetric
energy-momentum tensor of matter and $y^i$ denote, as usual,
Riemann normal coordinates.

The expression (66) gives us {\it total angular momentum
densities}, orbital and spinorial ones because the dynamical
energy-momentum tensor for matter $T^{ik} = T^{ki}$ is obtained
from the canonical one by means of the {\it Belinfante-Rosenfeld}
symmetrization procedure, and, therefore, includes material spin
\cite{Ber}.

Note that the normal coordinates $y^i$ form the components of the
local {\it radius vector} $\vec y$ with respect to the origin {\bf
P}. \footnote{A global radius vector does not exist in {\bf GR}.}

In consequence, the components of the $_m M^{ikl}(y)$ form a
local tensor density.

For the gravitational field we favourize and take the expression
proposed by Bergmann and Thomson \cite{Ber} as the gravitational
angular momentum pseudotensor
\begin{equation}
_g M^{ikl}(y) = _F U^{i[kl]}(y) - _F U^{k[il]}(y) + \sqrt{\vert
g\vert}\bigl(y^i{}_{BT} t^{kl} - y^k{}{}_{BT} t^{il}\bigr),
\end{equation}
where
\begin{equation}
_F U^{i[kl]}:= g^{im}{} _F U_m^{~[kl]}
\end{equation}
are the Freud superpotentials with the first index raised and
\begin{equation}
_{BT} t^{kl} := g^{ki}{}_E t_i^{~l} +
{g^{mk}_{~~,p}\over\sqrt{\vert g\vert}}{} _F U_m^{~[lp]}
\end{equation}
is {\it Bergmann-Thomson} gravitational energy-momentum
pseudotensor.

One can easily see that the expressions (66)-(67) are exactly the
material ($_m M^{ikl}$) and the gravitational ($_g M^{ikl}$) parts
of the canonical angular momentum complex (15).

The expressions (66)-(67) are  most closely related to the Einstein
canonical energy-momentum complex $_E K_i^{~k}$.
That is why we have applied them here and called them {\it
canonical}.

One can interpret the Bergmann-Thomson gravitational angular
pseudotensor as the sum of the {\it spinorial part}
\begin{equation}
S^{ikl} := _F U^{i[kl]} - _F U^{k[il]}
\end{equation}
and the {\it orbital part}
\begin{equation}
O^{ikl} := \sqrt{\vert g\vert}\bigl(y^i{}_{BT} t^{kl} - y^k{}_{BT}
t^{il}\bigr)
\end{equation}
of the gravitational angular momentum ``densities''.

Substitution of (66) and (67) (expanded up to third order) into
(63) gives the {\it canonical angular supermomentum tensors} for
matter and gravitation respectively
\begin{equation}
_m S^{abc}(P;v^l) = 2\bigl[\bigl(2v^a v^p -g^{ap}\bigr)\nabla_p
T^{bc} - \bigl(2v^b v^p - g^{bp}\bigr) \nabla_p T^{ac}\bigr],
\end{equation}
\begin{eqnarray}
_g S^{abc}(P;v^l)&=& \alpha\bigl(2v^p v^t -
g^{pt}\bigr)\bigl[\bigl(g^{ac}{} g^{br} - g^{bc}{}g^{ar}\bigr)
\nabla_{(t} R_{pr)}\nonumber\\
&+& 2 g^{ar}\nabla_{(t} {R^{(b}_{~~p}}^{c)}{}_{~~r)} - 2
g^{br}\nabla_{(t}{R^{(a}_{~~p}}^{c)}_{~~r)}\nonumber\\
&+& {2\over 3} g^{bc}\bigl(\nabla_r {R^r_{~(t}}^a_{~p)}
-\nabla_{(p}R^a_{~t)}\bigr) - {2\over 3} g^{ac}\bigl(\nabla_r
{R^r_{~(t}}^b_{~p)} - \nabla_{(p} R^b_{~t)}\bigr)\bigr].
\end{eqnarray}

Both these tensors are antisymmetric in the first two indices.

In vacuum, the gravitational canonical angular supermomentum
tensor (73) simplifies to
\begin{equation}
_g S^{abc}(P;v^l) = 2\alpha\bigl(2v^p v^t -
g^{pt}\bigr)\bigl[g^{ar} \nabla_{(p}{}{R^{(b}_{~~(t}}^{c)}_{~~r)}-
g^{br} \nabla_{(p} {R^{(a}_{~~t}}^{c)}_{~~r)}\bigr].
\end{equation}

Note that the orbital part
\begin{equation}
O^{ikl} = \sqrt{\vert g\vert}\bigl(y^i _{BT} t^{kl} - y^k _{BT}
t^{il}\bigr)
\end{equation}
{\it gives no contribution} to $_g S^{abc}(P;v^l)$. Only the
{\it spinorial part}
\begin{equation}
S^{ikl} = _F U^{i[kl]} - _F U^{k[il]}
\end{equation}
contributes.

Also, the canonical angular supermomentum tensors $_g S^{abc}(P;v^l)$
and  $_m S^{abc}(P;v^l)$ needn't any radius
vector too their own existing.\footnote{This is very good property because any global
radius vector does not exist in  {\bf GR}.}

Some final remarks:
\begin{enumerate}
\item In vacuum the quadratic form $_g S_a^{~b} v^a v_b$, where $v^a v_a =
1$, {\it is positive-definite}. This form gives the gravitational {\it superenergy
density} $\epsilon_g$ for a fiducial observer {\bf O}.
\item In general, the canonical superenergy and angular supermomentum tensors are uniquely
determined only along the world line of an observer {\bf O}. But
in special cases, e.g., in Schwarzschild spacetime or in Friedmann
universes, when there exists a physically and geometrically
distinguished four-velocity field, $v^i(x)$, one can introduce, in
an unique way, unambiguous fields $_g S_i^{~k}(x;v^l)$ and $_m
S_i^{~k}(x;v^l)$.

If we assume that the spacetime is {\it globally hyperbolic}, then
there also exists a distiguished, global timelike  vector field ${\vec v} : g(v,v) =
1$. One can use this vector field to global construction of the
canonical superenergy and angular supermomentum tensor fields.
\item It can be shown that the superenergy densities $\epsilon_g,
~~\epsilon_m$, which have dimension ${Joul\over metre^{5}}$,
exactly correspond to the Appel's {\it energy of acceleration} ${1\over 2}{\vec a}{\vec
a}$. The Appel's energy of acceleration plays fundamental role in Appel's approach
to classical mechanics.\cite{Appel} We have already told about that in main
text.
\item Recently we have noticed that the total superenergy
density is positive-definite or null for known gravitationally stable solutions to
the Einstein equations and negative-definite for gravitationally unstable
solutions. We have used this fact to study gravitational
stability, i.e., stability with respect small metric
perturbations, of many very known solutions to the Einstein equations. \cite{Gar1}
\item By using canonical gravitational superenergy and angular supermomentum
tensors one can prove that the exact gravitational waves carry
energy- momentum and angular momentum. \cite{Gar1}
\end{enumerate}

\acknowledgments

This paper was mainly supported by Institute of Mathematics, University of Szczecin
(Grant No 503-4000-230 351).

\end{document}